\documentclass[floatfix,twocolumn,nofootinbib,superscriptaddress,a4paper,preprintnumbers]{revtex4-1}
\usepackage{hyperref}
\usepackage{amsmath}
\usepackage{amssymb}
\usepackage{graphicx}
\usepackage[T1]{fontenc}
\usepackage{slashed}
\usepackage[utf8]{inputenc}
\usepackage{xspace}
\usepackage{color}
\usepackage{ulem}
\usepackage{array}

\newcommand{\calO}{{\cal O}}

\begin{document}


\title{Consistent origin of neutrino mass and freeze-in dark matter in large N theories}

\author{Zhi-Long Han}
\email{sps_hanzl@ujn.edu.cn}
\affiliation{Department of Physics, University of Jinan, Jinan 250022, P. R. China}

\author{Bin Zhu}
\email{zhubin@mail.nankai.edu.cn}
\affiliation{Department of Physics, Yantai University, Yantai 264005, P. R. China}

\author{Ligong Bian}
\email{lgbycl@cqu.edu.cn}
\affiliation{Department of Physics, Chongqing University, Chongqing 401331, P. R. China}

\author{Ran Ding}
\email{dingran@mail.nankai.edu.cn}
\affiliation{Center for High-Energy
Physics, Peking University, Beijing 100871, P. R. China}

\begin{abstract}
Most of what we concern in beyond standard phenomenology are the existence of tiny numbers. The well-defined principle for protecting the tiny number to be large from quantum correction is supersymmetry. However, such a nice framework is challenged by the non-observation of superpartners at LHC. Instead, we propose a new principle to realize a natural framework to explain the smallness of feebly interaction dark matter coupling and neutrino mass. The scalar sector as well as gauge sector must be extended to include $N$ copies as a price. It is found in this paper that the yukawa coupling $y$ as well as quartic coupling $\lambda$ scales with inverse power of $N$ to maintain perturbativity. In terms of the scaling behavior of couplings, the freeze-in dark matter becomes compatible with neutrino mass requirement. The biggest observation is that $y$ has to be evaluated by $1/N^{3/2}$ in type-I seesaw mechanism in order to obtain a large $N$ suppressed neutrino mass.  The intrinsic hierarchy between $1/\sqrt{N}$ and $1/N^{3/2}$ for yukawa coupling $y$ can be improved if we focus on the loop generated neutrino mass which can be relaxed to be $1/N$. The underlying reason for not use $1/\sqrt{N}$ is that freeze-in dark matter provides a lower bound for the scaling. Therefore the only choice of scaling for yukawa coupling is left to be $1/N$. Based on this simple scaling, we realiza an unified framework for explaining FIMP and neutrino mass.
\end{abstract}

\maketitle

\section{Introduction}
Physics Beyond the Standard Model (BSM)~\cite{Baer:2006rs,Dine:2015xga,Csaki:2016kln} is generally motivated by a series of tiny parameters, including the hierarchy problem of higgs mass~\cite{Feng:2013pwa}, non-zero neutrino mass~\cite{deGouvea:2004gd} as well as the $\theta$-term in strong CP problem~\cite{Dine:2000cj}. In addition, the existence of non-baryonic dark matter~\cite{Bertone:2004pz} also calls for BSM. Inspired by WIMP miracle and hierarchy problem, The Minimal Supersymmetric Standard Model (MSSM) is especially attractive. In this framework, the "smallness" (electroweak scale) of higgs mass is automatically protected by supersymmetry~\cite{Martin:1997ns}. Moreover, the correct relic abundance of Dark Matter (DM) candidate, neutralino, can be realized via thermal freeze-out mechanism through typical electroweak interaction~\cite{Jungman:1995df}. Thus the small number problems is solved in supersymmetric model naturally. However, such a nice framework is faced with severe challenges due to the null results of LHC direct searches and dark matter direct detections~\cite{Cui:2017nnn,Akerib:2016vxi,Aprile:2018dbl}. That strongly motivates us to consider alternatives of DM candidates among which Feebly Interaction Massive Particle (FIMP)~\cite{Hall:2009bx,Shakya:2015xnx,
Molinaro:2014lfa,Klasen:2013ypa,Elahi:2014fsa,Biswas:2016bfo,Bernal:2017kxu,
Bian:2018mkl,Bian:2018bxr} is favored.

The crucial feature of FIMP is that it always does not involve in the SM thermal bath, whose relic abundance is obtained by decay or scattering of particles with SM bath. which essentially requires a tiny coupling $y$ between SM and FIMP sectors to avoid thermal equilibrium. This then introduces additional tiny number~\cite{Cohen:2018cnq} from the point view of model building, i.e., new hierarchy problem of $y$ appears!
On the other hand, the tiny neutrino masse is accounted for seesaw mechanism through which ultra-light neutrino can be obtained via either small coupling or very massive right-handed neutrinos. It seems that FIMP and neutrino masses can be simultaneously explained by an unified small coupling $y$ in type-I like seesaw model~\cite{Asaka:2005an},

\begin{align}
\mathcal{L}=y L\phi N_R+M N_R N_R+m_{\phi}\phi^2+\lambda\phi^4,
\end{align}

where $\phi$ is $SU(2)_L$ doublet scalar. It is usually chosen to be SM higgs particle. There is also one possibility that $\phi$ is singlet under gauge group~\cite{Boehm:2006mi}. However the explicit gauge invariance violation must be treated as low energy effective theory so that it can be embedded into $SU(2)$ model as UV completion. This simple framework potentially causes three dangerous problems:

\begin{itemize}
\item Is neutrino mass consistent with FIMP dark matter requirement?
That is to say we wonder whether or not we can use single coupling to generate both FIMP and neutrino masses. That is greatly different from conventional treatment~\cite{Shakya:2015xnx,Molinaro:2014lfa,Drewes:2015eoa}. The discrepancy usually comes from the fact that FIMP dark matter requires the out-of-equilibrium behavior throughout the dark matter evolution in universe. That gives rise to a upper limit of the tiny coupling. Neutrino mass on the other hand puts a lower bound to the coupling $y$. Naively these two bounds have no overlap thus destroy the whole story. It will be proven to be wrong in our subsequent calculation.
\item Is the tiny coupling $y$ itself natural?
The tiny coupling $y$ has intrinsic hierarchy problem~\cite{Cohen:2018cnq} that need to be explained. Since there is no underlying symmetry protection for the lagrangian, the natural value of $y$ should be $\calO(1)$, and similar for scalar quartic coupling $\lambda$ .

\item The most important hierarchy problem is quadratic divergence to scalar $\phi$ mass.
As is demonstrated before, $\phi$ is doublet under $SU(2)$, the quadratic divergence are induced by Yukawa and gauge coupling as well as its self-interaction. The general form of quadratic divergence from one-loop diagram has the following type~\cite{Nilles:1982ik},
\begin{align}
\delta m^2\sim (y^2-g^2+\lambda^2)\frac{\Lambda^2}{16\pi^2}
\end{align}
such a $\Lambda^2$ sensitivity on UV scale leads to the gauge hierarchy problem when $\phi$ has mass much smaller than Plank scale. It is possible that ultra-tiny $y$ and $g$ could alleviate  fine-tuning problem of $\phi$. Therefore various aspects of this model is found to be related with each other by tiny couplings.
\end{itemize}

Our purpose is to provide a simple framework to answer above three questions. Motivated by the freeze-in hierarchy problem~\cite{Cohen:2018cnq}, the tiny couplings are easily to obtain by extending scalar $\phi$ to $N$-scalar sectors. Notice that the extension of fermion sector~\cite{Cohen:2018cnq} is only helpful for FIMP sector but not for neutrino mass and higgs hierarchy problem.

From a perspective of t'Hooft counting, $y^2 N$ is the actual coupling at large N limit. In order to retain in perturbative regime, $y^2N$ should be smaller than 1. Equivalently $y$ should be scaled with $1/\sqrt{N}$ which is the main starting point of our paper. Meanwhile the gauge coupling can be also regarded as $SU(N)$ gauge theory in large N limit, the gauge coupling is also evaluated at $1/\sqrt{N}$. The fine-tuning problem of $\phi$ is now improved. Based on that behavior, we manage to answer the three questions in a unified N-scalar framework.

The rest of this paper is layout as follows: section-\ref{sec:Nscalar} gives an overview of our model. In particular, the large N limit is put by hand in order to obtain ultra-tiny coupling. We are not here consider the UV completion of this model and just give a benchmark model explicitly. Furthermore the improvement of fine-tuning is explained in detail. In section-\ref{sec:Neutrino}, we consider whether or not the benchmark model can give a consistent origin of neutrino mass and FIMP dark matter. The scaling behavior of $y$ must be further relaxed into $1/N^{3/2}$ so that the two issues are compatible with each other. Such a unnatural scaling can be improved when we consider two-loop neutrino mass generation mechanism.

\section{Large N Field Theory For Naturally tiny coupling and hierarchy problem of scalar}
\label{sec:Nscalar}
In this section, we propose a benchmark model on the consistent origin of freeze-in dark matter and neutrino mass. The lagrangian of N-sector is given with $\phi_\alpha$ being SM doublet

\begin{align}
\mathcal{L}=y^{ij}L_i\sum_{\alpha=1}^{N}\phi_{\alpha}N_j + M_N N_j^c N_j
+\sum_{\alpha=1}^{N}m_\alpha^2|\phi_{\alpha}|^2+\lambda|\phi_{\alpha}|^4
\end{align}

where $i,j$ indices stand for the three generation leptons. $L_i$ is left handed leptons while $N_j$ is right-handed neutrino~\cite{Fritzsch:1974nn,Yanagida:1980xy}. $\alpha$ accounts for N scalar fields in the lagrangian~\cite{Arkani-Hamed:2016rle}. In general, the keV scale right-handed neutrino, i.e. sterile neutrino can generate sizable relic density via Dodelson-Widrow (DW) mechanism~\cite{Dodelson:1993je}. However such a mechanism is disfavored by combined constraints of X-ray and Lyman-$\alpha$ forest~\cite{Seljak:2006qw}. In addition, the very light sterile neutrino is not compatible with neutrino mass in type-I seesaw mechanism where $10^{14}$ GeV right hand neutrino is favored for order $1$ coupling $y$:

\begin{align}
m_{\nu}=\frac{y^2 v_{\phi}^2}{M}\sim 0.1 \text{eV}
\end{align}

Alternatively, freeze-in mechanism become attractive to host sterile neutrino DM. As a consequence, tiny $y$ is not only suitable for freeze-in dark matter but compatible with neutrino mass. The natural origin of tiny yukawa is derived by the minimal N scaling of coupling to maintain perturbativity. This approach has been well developed in 'tHooft large N expansion~\cite{tHooft:1973alw} and called  'tHooft counting for simplicity.

It is easy to find that each feynman diagram associated with right-hand neutrino contains a factor $r$,

\begin{align}
r\sim (y^2N)^{V/2}N^{-E/2}\,.
\end{align}

The actual physical coupling, i.e., t'Hooft coupling is now $y^2 N$ rather than $y^2$. In order to maintain perturbativity, the Yukawa coupling must be smaller than $1/\sqrt{N}$. This scaling behavior thus provides a natural method for capturing a tiny coupling. Follow this logic, the self-interacting coupling $\lambda$ in scalar sector should be scaled as $1/N$ which is the main spirit of $N$-inflation~\cite{Dimopoulos:2005ac}. Then it allows us to testify the validity of this model to explain neutrino mass, relic density  and scalar hierarchy simultaneously. Before that, the scalar hierarchy is the most dangerous problem of our model that we must cope with. From a perspective of black hole entropy argument in large degree of freedom, the effective cut-off can be naturally reduced to be around TeV scale. The large $N$ sector is introduced to soften hierarchy problem of scalar particle~\cite{Arkani-Hamed:2016rle} where statistical distribution of N sector favors a naturally light higgs. Another method to solve hierarchy problem is also based on large N new degrees of freedom. The corresponding graviton-graviton scattering at energy $E$ are enhanced by $N$. As a consequence, the ultimate UV cutoff is soften by~\cite{Dvali:2007hz,Dvali:2007iv,Dvali:2011aa}

\begin{align}
\Lambda_{UV}\sim 4\pi\frac{M_{P}}{\sqrt{N}}\,.
\end{align}

The price that we paid is the additional huge number of degrees of freedom. Here we propose alternative consideration on hierarchy problem of scalar particle which is equivalent to black hole argement but has direct relationship with loop correction. Thanks to the tiny couplings~\cite{Dimopoulos:2005ac} in this mode, the corresponding one-loop correction is suppressed by large $N$. Only the yukawa coupling scales with $1/\sqrt{N}$, it provides the dominated contribution in the loop correction compared with other couplings,

\begin{align}
\delta m^2\sim \frac{\Lambda^2}{16\pi^2 N},
\label{eqn:loop}
\end{align}

This method is similar with the solution of $\eta$ problem~\cite{Ashoorioon:2011aa} in inflation. The fine-tuning can be quantified by the simple low energy measure $\delta m_{\phi}^2/m_{\phi}^2$~\cite{Baer:2013gva}. Due to large N suppression in equation~(\ref{eqn:loop}), the quadratic divergence is soften naturally. If we further allow the yukawa coupling to scale with $1/N$, the fine-tuning becomes much smaller than that in $1/\sqrt{N}$. The essential idea is that the scalar potential is not a single field but a collection of $N$ fields.

\begin{figure} [htbp]
\begin{center}
\includegraphics[width=0.4\textwidth]{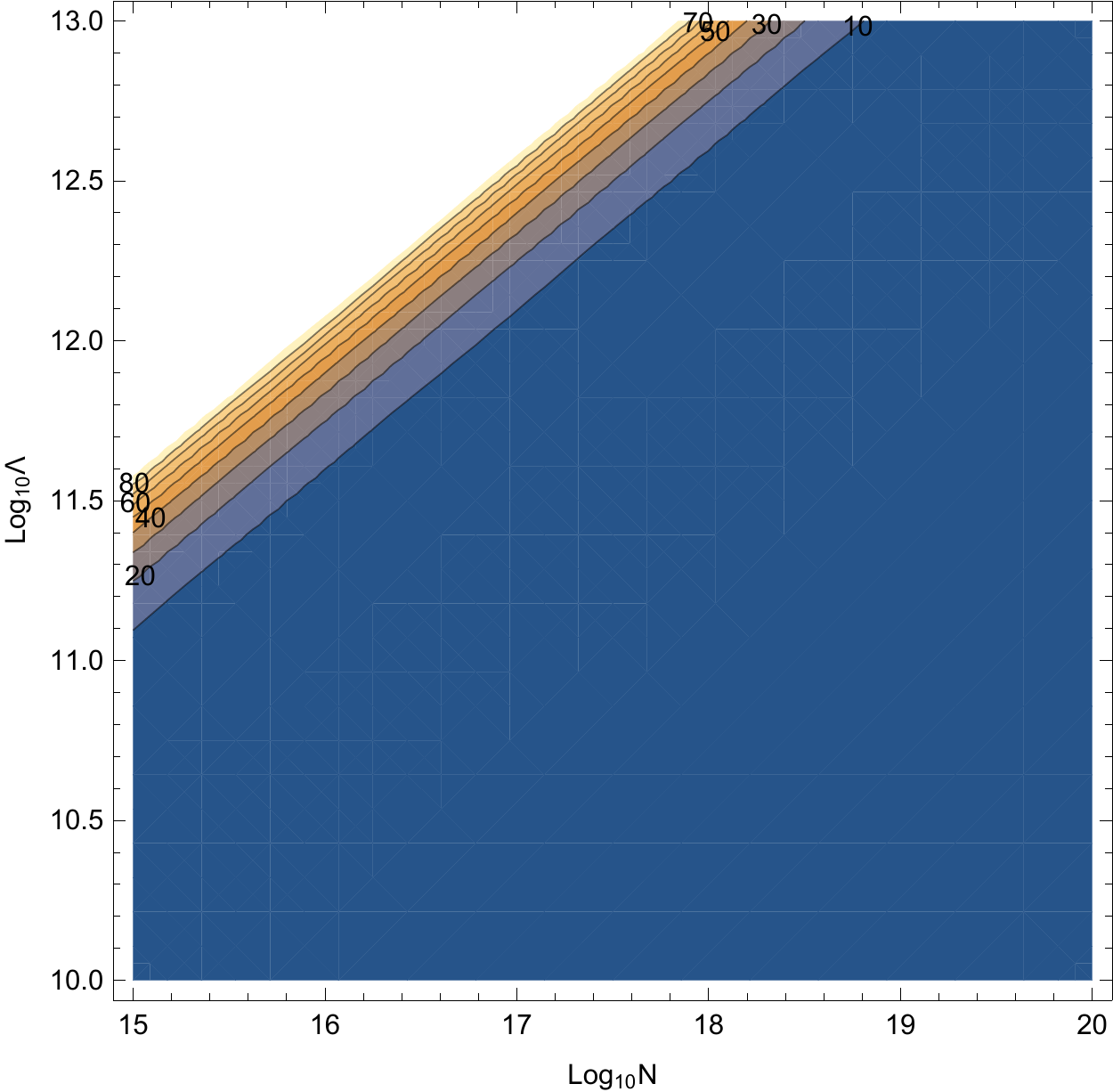}
\includegraphics[width=0.4\textwidth]{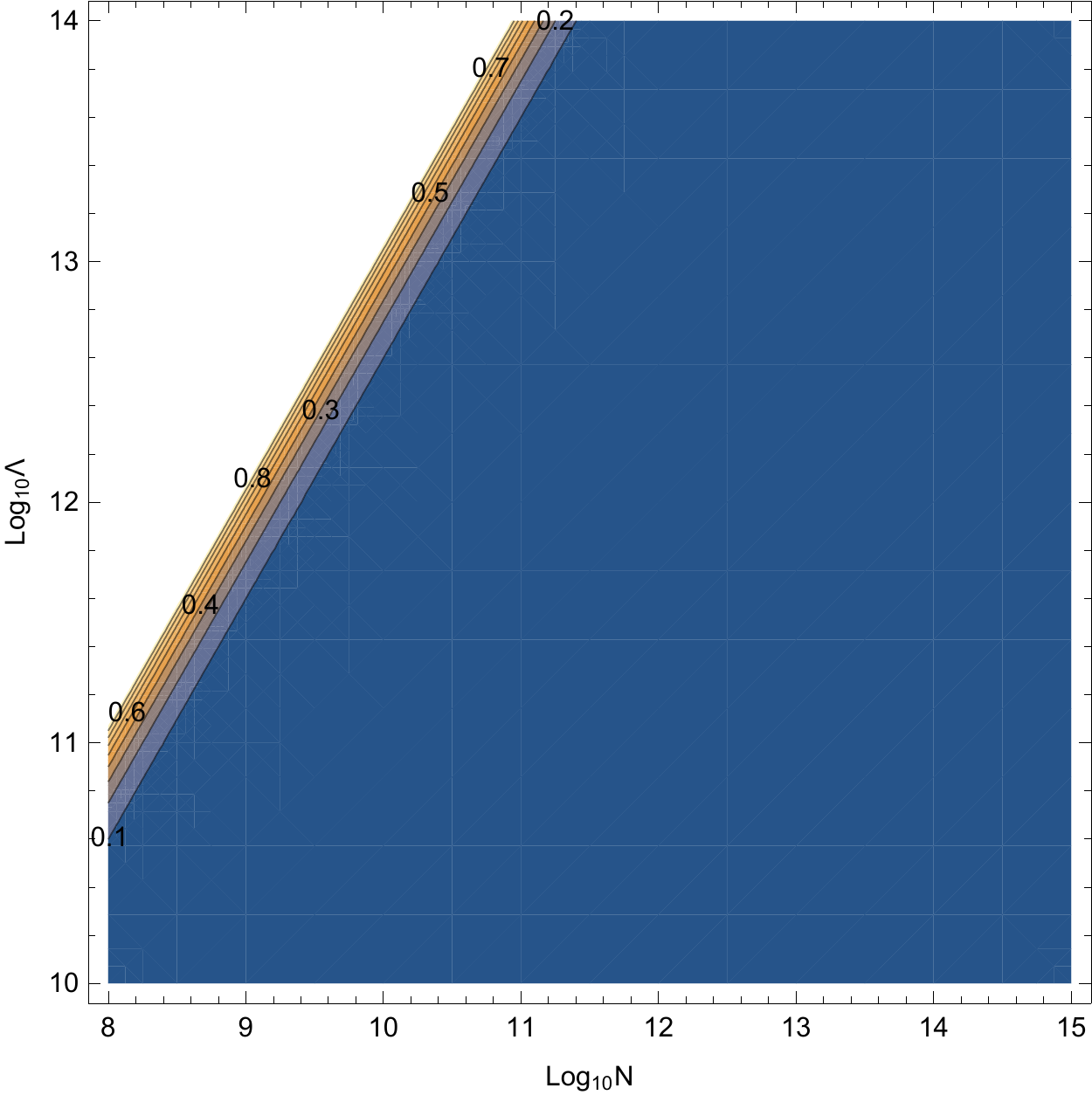}
\end{center}
\caption{The dependence of fine-tuning on $\log N$ and $\log\Lambda$ is given explicitly with $m_{\phi}=100~\text{GeV}$.}
\label{fig:finetuning}
\end{figure}

In figure~\ref{fig:finetuning}, we plot the behavior of fine-tuning measure as a function of $N$ and $\Lambda$ with $m_{\phi}$ setting to be $100$ GeV.  The upper panel of  figure~\ref{fig:finetuning} indicates that $N$ must be larger than $10^{15}$ in order to make fine-tuning under control when the yukawa coupling scales with $1/\sqrt{N}$.  On the opposite, if we choose $y$ to scale with $1/N$, it requires much smaller number of $N$.

So the fine-tuning becomes moderate when large $N$ is taken. Of course, much larger $N$ continues to improve the fine-tuning. But quite a larger number of $N$ might be tension with FIMP and neutrino mass which will be considered subsequently. From now on, we take a benchmark value of fine-tuning throughout the paper $\Delta=1$ The cut-off scale varies with $4\pi m_{\phi} \sqrt{N}$ and $4\pi m_{\phi} N$ for $1/\sqrt{N}$ and $1/N$ respectively.

\section{Neutrino mass generation and relic density}
\label{sec:Neutrino}

In type-I seesaw, the $0.1$ eV neutrino mass mass requires $10^{14}$ GeV sterile neutrinos for order $1$ Yukawa coupling.  Therefore the dangerous hierarchy problem is solved by ultra-heavy sterile neutrino. Because fermion mass is always technically natural, we do not worry about neutrino mass any longer. It seems there is no need to consider large-N sector. However it is easy to find that we lost the possibility of sterile neutrino DM as a price.

Here we insist sterile neutrino to be FIMP dark matter i.e. $m_{\phi}>m_{N}$. The coupling $y$ must be very tiny which re-introduces the hierarchy problem for both FIMP and neutrino mass. Furthermore people thought that  neutrino mass is not consistent with FIMP. It can be formulated as a two-scale problem~\cite{Molinaro:2014lfa}:

\begin{align}
y_{\text{FIMP}}<10^{-8}, \; \; y_{\text{neutrino}}>10^{-6}\,.
\end{align}

That motivates people to consider them individually. For example we only treat the first generation sterile neutrino as DM with the other two as seesaw mechanism. This un-natural behavior can be corrected once we consider the neutrino mass seriously. We assume both of them are consistent with each other which is the essential point of this paper. As we know the tHooft counting constrains $y$ to scale with $1/\sqrt{N}$. The corresponding neutrino mass is thus,

\begin{align}
m_{\nu}=\frac{N y^2 v_{\phi}^2}{M_N}=\frac{v_{\phi}^2}{M_N}\,,
\end{align}

here the existence of $N$ means there are $N$ scalars contributing to neutrino mass. Naively, the $v_{\phi}$ is regarded as a free parameter so that neutrino mass is easy to obtain by taking a almost zero vacuum expectation value. In fact $v_{\phi}$ is not free parameter which must be given through the effective potential of $\phi$,

\begin{align}
v_{\phi}=\frac{m_{\phi}}{\sqrt{\lambda}}
\end{align}

Since $\lambda$ scales as $1/N$, it does not produce any suppression but enhancement i.e., $m_{\nu}\sim m_{\phi}^2 N/4 m_N$. Under this condition, large N sector does not solve the neutrino mass problem unless we take a further constrained scaling such as $y\sim 1/N^{3/2}$. The corresponding neutrino mass in this scaling reads,

\begin{align}
m_{\nu}=\frac{m_{\phi}^2}{4 m_N N}\,,
\end{align}

It is easy to see that only this scaling can generate reliable neutrino mass.  For now they are only independent parameters. Together with the relic density requirement, the three parameters are reduced to two: $m_{\phi}$ and $m_N$. The relic density of FIMP sterile neutrino is determined by the Boltzman equation \cite{Hall:2009bx},

\begin{align}
\frac{dn_{\alpha}}{dt}+3H n_{\alpha}=\frac{g_{SM}m_{SM}^2}{2\pi^2}T K_1\left[\frac{m_{SM}}{T}\right]\Gamma[\phi_{\alpha}\rightarrow L_i N_j]
\end{align}

The resulting relic density is
\begin{align}
\Omega h^2=\frac{1.09\times 10^{27}g_{\phi}}{g_{\star}^s\sqrt{g_{\star}^{\rho}}}
\frac{m_N N \Gamma}{m_{\phi}^2}.
\end{align}

The requirement of $\Omega h^2\sim 0.12$ \cite{Aghanim:2018eyx} further fixes $N$ in terms of $m_{\phi}$ and $m_N$,
\begin{align}
N=6.6\times 10^{12}\sqrt{m_N/m_{\phi}}.
\label{eqn:numberN}
\end{align}

\begin{figure} [htbp]
\begin{center}
\includegraphics[width=0.4\textwidth]{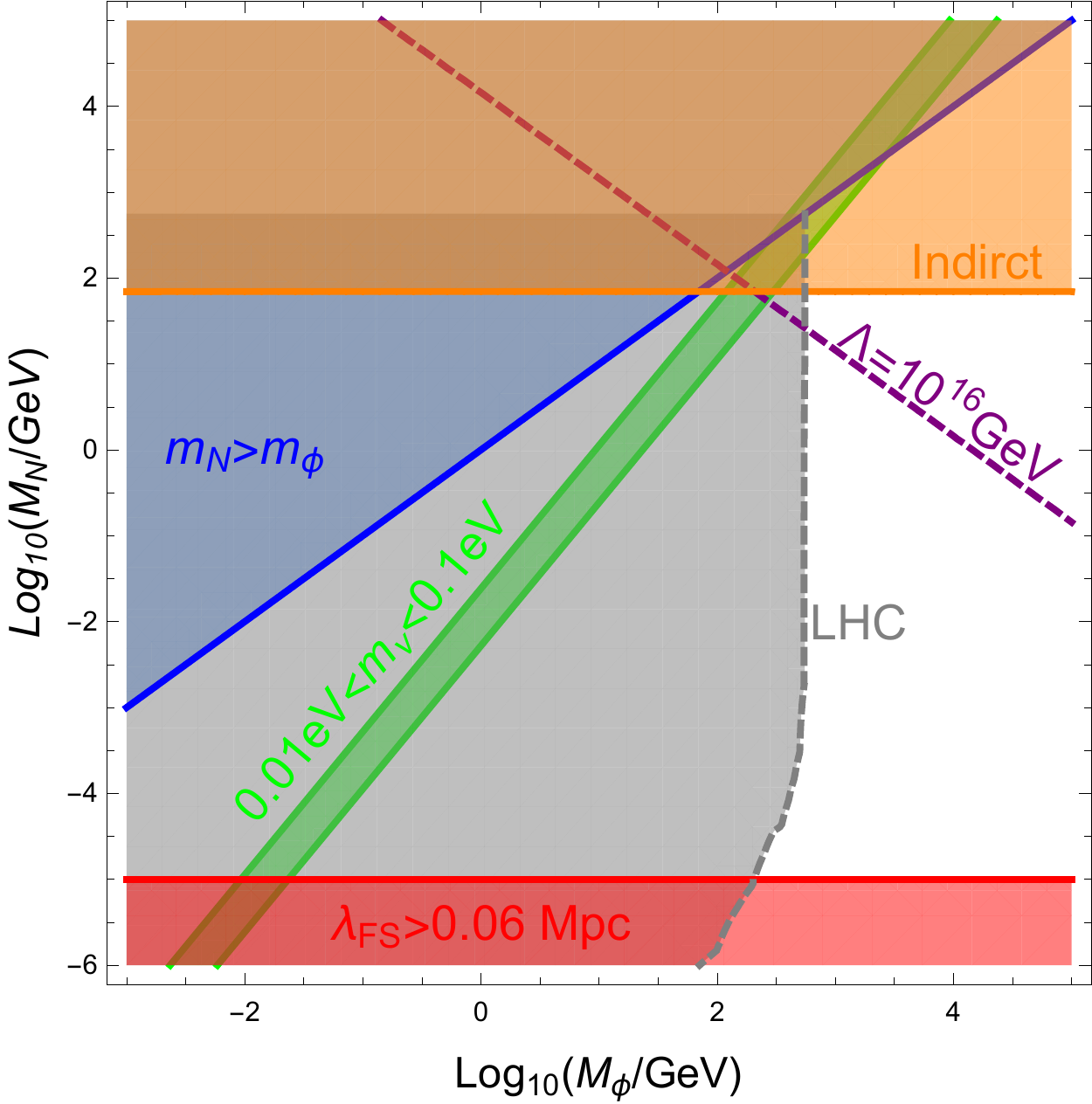}
\end{center}
\caption{The dependence of correct relic abundance and neutrino mass in $[\log_{10}m_{\phi},~\log_{10}m_N]$ plane (green band). In addition, the requirement for $\Gamma<H$ is trivial to satisfy.}
\label{fig:neutrino}
\end{figure}

In terms of equation~(\ref{eqn:numberN}), the neutrino mass becomes a function of $m_{\phi}$ and $m_{N}$. In figure~\ref{fig:neutrino}, we show the correct relic abundance and neutrino mass favoured region varying along $m_{\phi}$ and $m_{N}$. There are several constraints that this contour must satisfy:

\begin{itemize}
\item The FIMP relic density must be smaller than $0.1$. In this plot, each contour are required to satisfy $\Omega h^2=0.12$.
\item To interpret neutrino oscillation, one typically needs $0.01~\text{eV}<m_\nu<0.1$ eV \cite{Esteban:2018azc}. Then combining the requirements for $\Omega h^2$ and $m_\nu$, the favoured region in $m_\phi-m_N$ is determined.
\item FIMP dark matter requires quite tiny coupling in order to departure from thermal equilibrium, which is schematically written as
\begin{align}
\Gamma[\phi\rightarrow L_iN_j]<H_{\phi}
\end{align}
This constraint is highly non-trivial, and gives a upper bound of the coupling $y<5.9\times10^{-9} \sqrt{m_{\phi}/\text{GeV}}$. However since $y$ scales with $1/N^{3/2}$, it can satisfy the bound easily.
\item The cut-off scale should be larger than GUT scale otherwise the solution to hierarchy problem is not attractive. Here we should mention that, since $y=1/N^{3/2}$, the dominate fine-tuning contribution comes from $\lambda=1/N$. Thus the physical cut-off is now $4\pi m_{\phi}N$.
    From figure~\ref{fig:neutrino}, it is clear that $\Lambda\sim10^{16}$ GeV is possible with $m_N\lesssim m_\phi\sim10^2$ GeV.
\item The free-streaming length characterises the structure formation, which places a lower bound on DM mass. It is estimated by \cite{Adulpravitchai:2015mna}
    \begin{equation}
    \lambda_\text{FS}\approx 0.047 ~\text{Mpc} \left(\frac{10 \text{keV}}{m_N}\right).
    \end{equation}
    Observations from Lyman-$\alpha$ forest \cite{Irsic:2017ixq} have excluded
    $\lambda_\text{FS}\gtrsim 0.06$ Mpc. Therefore, $m_N\gtrsim10$ keV is required.
\item Due to the doublet nature of $\phi$, its charge component $\phi^\pm$ can be pair produced at LHC. As the Yukawa coupling $y$ is tiny, the decay $\phi^\pm\to N\ell^\pm$ will lead to long-lived charged particle signature. We consider the exclusion region obtained by Ref.~\cite{Hessler:2016kwm}, where $m_{\phi^\pm}\lesssim 560$ GeV is excluded. In the plot, we take degenerate mass spectrum of $\phi$ for simplicity. It is quite clear that most of the favoured region is excluded by the LHC search.
\item Since the sterile neutrino $N$ mixes with light neutrino $\nu$, the decay products of $N$ would lead to observable $\gamma-$ or $X-$ray signatures. For heavy $N$ near or above EW scale, a rough reinterpretation of the decay channels $N\to W^{(*)} l,Z^{(*)}\nu, h^{(*)}\nu$ by Ref~\cite{Cohen:2016uyg,Cohen:2018cnq} indicates that $m_N\gtrsim70$ GeV is excluded. In this way, the final corner uncovered by LHC search is also eliminated.
\end{itemize}

In summary, the naive combined constraints from LHC and indirect detection have already excluded all the favoured parameter space for relic abundance and neutrino mass. Notably, the LHC bound on $m_\phi$ can be weakened down to about 160 GeV when $\phi^\pm$ decay dominantly into next-to-lightest odd particle.
In this way, there is still a corner to survive all constraints.

Throughout the calculation of neutrino mass in type-I seesaw model, we found that the intrinsic problem comes from the discrepancy between large N requirement  $1/\sqrt{N}$ and neutrino mass requirement $1/N^{3/2}$. Therefore we are left with an intrinsic fine-tuning $N$ which is comparable with eletroweak hierarchy problem. It strongly motivates us to consider higher loop neutrino mass generation mechanism~\cite{Ma:2006km,Sierra:2014rxa,Nishiwaki:2015iqa,Cai:2017jrq,
Kashiwase:2015pra,Ding:2016wbd,Simoes:2017kqb,Guo:2017gxp,Ding:2018jdk,
Han:2018zcn} where higher powers of $y$ is possible. As we know the generation of neutrino masses via quantum correction is a viable scenario. The most famous and historic example is Zee-Babu model where double-charged scalar is included to generate neutrino mass via two-loops. The disadvantage of this model is the lack of dark matter candidate. We can start with simple one-loop neutrino mass mass. It is easy to identify that it can not provide large $N$ suppression to neutrino mass when $y=1/N$.

Based on the fact that the higher power of $y$, the lower dependence of $N$ is required. For example $(y^2/4\pi)^m$ determines the general form of neutrino mass loop effect. Here $m=2$ corresponds to one loop, $m=3$ corresponds to two-loop. The two-loop generated neutrino model has been classified in~\cite{Sierra:2014rxa}. Depending on different reps and field contents, there are two classes of models: seven-particle model and six particle model. Here we take the two-loop model proposed by in Ref.~\cite{Ma:2007gq} as an example, which is in general belongs to class 1.a with neutral fermions and singlets.

\begin{align}
\mathcal{L}\supset y^{ij} L_i \phi N_j + y_S^{ij} N_i N_j S  +\mu S^3
\end{align}

Here, $\phi$ and $N$ carry non-trivial charge under discrete symmetry. So the DM candidate $N$ is stable. Provided $y\sim y_S$,
the neutrino mass is then estimated as follows,

\begin{align}
m_{\nu}\sim \frac{\mu y^3}{4(2\pi)^8}
\end{align}

$\mu$ is dimensional parameter and proportional to $\lambda m_S$. Further taking $m_S\sim m_\phi$, the neutrino mass at scaling $y\sim1/N$ can be written as follows,

\begin{align}
m_{\nu}=\frac{m_{\phi}^3}{512\pi^4m_N^2 N}
\end{align}

\begin{figure}[htbp]
\begin{center}
\includegraphics[width=0.4\textwidth]{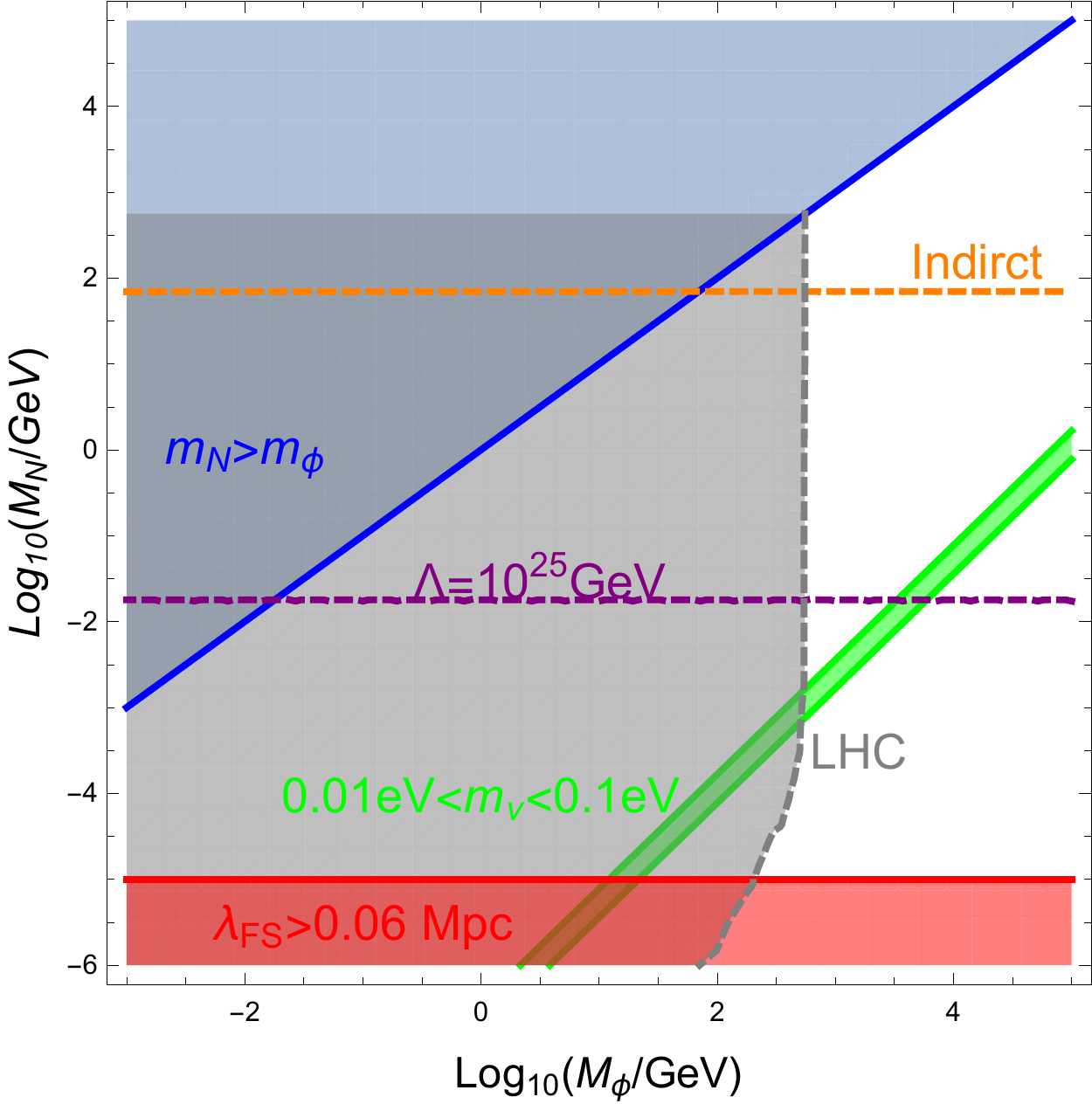}
\end{center}
\caption{Same as Fig.~\ref{fig:neutrino}, but for two-loop model. Because $N$ is stable in this case, the indirect search bound do not applicable here.}
\label{fig:radiative}
\end{figure}

Figure \ref{fig:radiative} shows the favoured region for correct relic abundance and neutrino mass as well as the corresponding bounds. It is obvious that $m_\phi\gtrsim 600$ GeV with $m_{N}\gtrsim 1$ MeV could satisfy all constraints.

It is similar with type-I seesaw where the scaling behavior of $y$ is now $1/N$ rather than $1/N^{3/2}$. Motivated by this logic, higher order loop generation can recover the $1/\sqrt{N}$ scaling. However $1/\sqrt{N}$ is contradict with FIMP dark matter, i.e. relic density without large N suppression. It is similar with neutrino mass at type-I seesaw, the relic density without large N suppression destroys the consistency between neutrino mass and FIMP dark matter. That is to say, relic density provides a lower bound of scaling $1/N$, neutrino mass on the other hand gives a upper bound scaling $1/N^{3/2}$.

\section{Conclusion}
Naturalness, neutrino mass and existence of dark matter require new physics beyond the Standard Model. The strong intrinsic connection among these three problems lead to unified model building such as supersymmetry. Unlike supersymmetry, we provide a simple framework to explain them where SM is extended into $N$ scalars sector. From perspective of tHooft counting, the coupling must be scaled around inverse power of $N$. As a consequence the hierarchy problem of scalars is improved. In type-I seesaw model, the freeze-in dark matter is compatible with neutrino mass with $y$ being $1/N^{3/2}$. It leads to another misalignment problem where perturbativity requires $y$ to be $1/\sqrt{N}$ while neutrino mass requires $y$ to be $1/N^{3/2}$. Under tight constraints from LHC and indirect detection, the minimal extension of type-I seesaw model is already excluded. However, if $\phi^\pm$ decay dominantly into next-to-lightest odd particle, a corner at $m_N\lesssim m_\phi\sim[160,250]$ GeV might be still possible. When we go further into two loop neutrino mass, the scaling is reduced to be $1/N$. Viable parameter space is $m_\phi\gtrsim 600$ GeV with $m_{N}\gtrsim 1$ MeV. Finally,  no matter what we do, the freeze-in dark matter forbids the scaling to be smaller than $1/N$. Therefore, only $1/N$ is possible to solve these three problems simultaneously.
\section*{Acknowledgements}
Bin Zhu is supported by the National Science Foundation of China (11747026  and 11805161) and Natural Science Foundation of Shandong Province under Grant No. ZR2018QA007. Zhi-Long Han is supported by National Natural Science Foundation of China under Grant  No. 11605075 and No. 11805081, Natural Science Foundation of Shandong Province under Grant No. ZR2018MA047, No. ZR2017JL006 and No. ZR2014AM016.

\end{document}